\documentclass[conference,compsoc]{IEEEtran}
\usepackage[utf8]{inputenc}
\usepackage[T1]{fontenc}
\usepackage{amsmath,amssymb}
\usepackage{booktabs}
\usepackage{xcolor}
\usepackage[hidelinks]{hyperref}
\usepackage{pgfplots}
\pgfplotsset{compat=1.16}
\newcommand{\gopt}{\gamma_{\mathrm{opt}}}

\IEEEoverridecommandlockouts
\begin{document}

\title{Similarity-Weighted Leiden-CPM Clustering of a 7.4-Million-Node Biomedical Citation Graph: Weight Alignment, Scale, and Sparsity}

\author{\IEEEauthorblockN{Hossein Mohasel Arjomandi\thanks{This work is not related to the author's current employment at Microsoft. The views expressed are the author's own. This work has been submitted to the IEEE for possible publication. Copyright may be transferred without notice, after which this version may no longer be accessible.}}
\IEEEauthorblockA{Microsoft, Redmond, WA, USA\\
hosseinm@microsoft.com}
}

\maketitle

\begin{abstract}
Clustering citation networks into research communities supports science mapping, research evaluation, and the analysis of funding portfolios.
Biomedical articles also carry metadata, such as MeSH terms, publication year, and abstracts, that can be turned into similarity-based edge weights for Leiden clustering under the Constant Potts Model (CPM).
Studies of hybrid relatedness compare the accuracy of such clusterings, but how the weights interact with the CPM resolution parameter on a large graph is less clear.
Here it is shown that, on a citation network derived from the cell biology research literature (7.4 million nodes, 57.5 million edges), the effect of the weights is governed mainly by their scale and by edges of weight zero, and that, once these are controlled, weighting gives modestly more coherent clusters.
Edges were weighted with seven topological and semantic similarity features and clustered with Leiden-CPM, using a method updated to fix a bug in how earlier code attached weights to edges.
Mean weights of 0.0015 to 0.23 raised the effective resolution 4- to 650-fold; dividing by the mean restored most of the coverage, except for topological weightings, in which the 47\% of edges with weight zero hold most of it. At equal coverage, weighted clusters were more coherent than unweighted ones, also on features outside the weight.
For weighted CPM clustering, weights should therefore be attached to edges by key, normalized or reported with their effective resolution, and checked for zero-weight edges.
\end{abstract}

\begin{IEEEkeywords}
citation networks, community detection, Leiden algorithm, Constant Potts Model, PubMed, edge weighting, reproducibility
\end{IEEEkeywords}

\section{Introduction}
Grouping publications into coherent research communities supports science mapping, research evaluation, and analysis of funding portfolios.
Many large-scale methods cluster the citation graph using its links alone.
Yet each biomedical article also carries metadata, such as MeSH indexing terms, publication year, and an abstract, that could help separate topically related citations from incidental ones.

This paper studies one way to use those metadata: assign every citation edge a similarity-based weight and cluster the weighted graph with Leiden optimizing the Constant Potts Model (CPM).
The weights leave the graph topology unchanged, so topological quality measures remain meaningful.
The findings concern anyone who clusters citation graphs with weighted Leiden-CPM, for example with \texttt{leidenalg} or igraph, for science mapping or portfolio analysis.
This paper is derived from the author's M.S. thesis~\cite{Arjomandi_2024}. The thesis reported that the weighting lowers coverage sharply and that rescaling the weights to $[1,10]$ keeps it near the unweighted baseline. With the method updated to fix a bug in how weights were attached to edges (Section~\ref{sec:align}), coverage is instead governed by weight scale and by zero-weight edges, and weighting makes clusters more coherent at equal coverage.
This paper makes three contributions.
\begin{itemize}
\item With repeated seeds, it separates weight scale from feature content by normalizing each weighting to mean one, and relates the scale to an effective resolution (Section~\ref{sec:scaletest}).
\item It identifies the edges without topological overlap as the main remaining cause of lost coverage (Section~\ref{sec:scaletest}).
\item It compares cluster coherence at equal coverage against an unweighted resolution sweep, and quantifies the gains (Section~\ref{sec:coherence}).
\end{itemize}
A companion paper~\cite{Arjomandi_A} describes the data pipeline and feature construction, which this paper does not repeat; the weighting (Section~\ref{sec:method}) is summarized from it.
Thesis results are restated as reported.

\section{Related Work}
\emph{Clustering algorithms.}
Leiden~\cite{Traag_2019} improves on Louvain by guaranteeing well-connected communities, and optimizes either modularity or CPM.
Modularity suffers from a resolution limit~\cite{Fortunato_2007}; CPM~\cite{Traag_2011} exposes the density threshold as a resolution parameter $\gamma$.
Infomap~\cite{Rosvall_2008} instead finds communities by compressing the description of a random walk; it is not used here.

\emph{Relatedness measures for publication clustering.}
Bibliographic coupling~\cite{Kessler_1963} and co-citation~\cite{Small_1973,Small_1985} are classical citation-based relatedness measures.
Waltman et al.~\cite{Waltman_2010} unify mapping and clustering of bibliometric networks, and a later methodology compares relatedness measures for clustering publications~\cite{Waltman_2020}.
Boyack et al.~\cite{Boyack_2011} compare text-based similarity approaches on more than two million biomedical publications.
Yun et al.~\cite{Yun_2020} cluster the literature with structural information only.

\emph{Hybrid relatedness.}
Waltman and van Eck~\cite{Waltman_2012} propose a methodology for constructing a publication-level classification system of science.
Boyack and Klavans~\cite{Boyack_2020} compare clusterings from direct citation, extended direct citation, a textual measure, and citation--text hybrids on nine million documents, and find that a hybrid of direct citation and PubMed related-article scores gives the most accurate clusters in aggregate.
Ahlgren et al.~\cite{Ahlgren_2020} enhance direct citations with bibliographic coupling, co-citation, extended direct citations, and the BM25 text measure, and evaluate the resulting clusterings against MeSH.
These studies evaluate which relatedness measure gives the most accurate clusters.
This paper asks a different question: how a weighted combination of seven features interacts with the CPM resolution parameter on a 7.4-million-node graph.
It does not compare against their accuracy evaluations.

\emph{Data.}
The network used here covers extracellular-vesicle research. Earlier work applies overlapping-community detection~\cite{Jakatdar_2022} and a core--periphery analysis~\cite{Wedell_2022} to networks of this literature.
Peel et al.~\cite{Peel_2017} caution that metadata need not align with network communities, a point taken up in Section~\ref{sec:discussion}.

\section{Data}
\label{sec:data}
The graph and its features are described in a companion paper~\cite{Arjomandi_A}; only what this study needs is summarized here.
Parsed PubMed metadata for the same period are public in the dataset ``Parsed Open Citations and PubMed Data''~\cite{Arjomandi_2024_data} (DOI~\texttt{10.13012/B2IDB-5216575\_V1}, CC~BY~4.0). That dataset also contains an Open Citations edge list, which this study does not use.
Article metadata (year, MeSH terms, title, abstract) come from the PubMed annual baseline, downloaded in January 2024 (about 34 million records by 2023)~\cite{Arjomandi_2024}.

The experiments use a curated citation network (CEN), derived from the cell biology literature and centered on extracellular vesicles (exosomes), with about 14 million nodes and 92 million edges~\cite{Arjomandi_2024}; its edge list, with integer node IDs, is public~\cite{Park_2023_data}.
Restricting it to articles indexed in PubMed that have a publication year, MeSH terms, a title, and an abstract gives the graph called CENM, with 7,445,898 nodes and 57,512,086 edges~\cite{Arjomandi_2024}.
The graph is fixed after this step, so weights never alter its topology.

\section{Method}
\label{sec:method}
\subsection{Edge features}
Each edge of CENM carries seven features $f_1,\dots,f_7$, defined in the companion paper~\cite{Arjomandi_A}.
Four are topological: bibliographic-coupling count ($f_1$) and Jaccard coefficient ($f_2$), and co-citation count ($f_3$) and Jaccard coefficient ($f_4$).
Three are semantic: a step function of the publication-year gap ($f_5$, equal to $1$ for a gap of at most five years), cosine similarity of MedCPT title--abstract embeddings~\cite{Jin_2023} ($f_6$; observed range $[0.498,1]$ and interquartile range $[0.751,0.827]$ on CENM~\cite{Arjomandi_A}, so $f_6$ adds a nearly constant term to the weight), and a MeSH-tree similarity ($f_7$).
Features $f_1$--$f_4$ and $f_7$ are min--max scaled to $[0,1]$ over the whole network; their minimum is $0$, so this equals the division by the maximum described in the thesis.
The companion paper's audit of the feature table~\cite{Arjomandi_A} found that $f_7$ is nonzero on only 0.28\% of the edges, so the MeSH weight in the experiments below has almost no effect; that paper recommends a mean or best-match aggregation instead of the median.
Features exist only for edges of the original graph.

\subsection{Edge weight}
With user weights $w_i\ge 0$, $\sum_i w_i=1$, the edge weight is
\begin{equation}
s(a,b)=\sum_{i=1}^{7} w_i f_i(a,b).
\end{equation}
Setting $w_5=w_6=w_7=0$ removes the semantic features, and an edge then has weight zero if it has no shared references and no co-citations. This holds for 47\% of the edges (Section~\ref{sec:thesisw}).

\subsection{Clustering and metrics}
\label{sec:metrics}
Leiden-CPM maximizes
\begin{equation}
H=\sum_{c}\left(e_c-\gamma\binom{n_c}{2}\right),
\end{equation}
where $e_c$ is the total edge weight inside cluster $c$, $n_c$ its number of nodes, and $\gamma$ the resolution~\cite{Traag_2011,Traag_2019}. Larger $\gamma$ demands denser clusters.
Leiden optimizes $H$ with the edge weights $s(a,b)$.
The reported \emph{CPM score} is different: following the thesis code, it evaluates $H$ for the resulting partition on the unweighted graph, so $e_c$ is the number of edges inside $c$, at $\gamma=0.05$, summed over non-singleton clusters.
The score is on the same scale for every weighting, but it is not neutral: the unweighted run at $\gamma=0.05$ optimizes exactly this score, so it favors the baseline, and coverage mostly tracks the resolution.
Three measures are reported: Cov1, the percentage of nodes in non-singleton clusters; Cov10, the percentage in clusters with more than ten nodes; and the CPM score.
Clusters of ten or fewer articles are unlikely to be meaningful research topics, so Cov10 is the more relevant coverage measure.
The new runs add three coherence measures over intra-cluster edges, which can reward weighting: the mean of $f_6$, the share of edges with $f_5=1$, and the mean of $f_2$. Over all edges of CENM they are 0.788, 0.450, and 0.0172.

\subsection{Scale equivalence}
\label{sec:scale}
Multiplying every edge weight by $\alpha>0$ multiplies $e_c$ by $\alpha$, so
\begin{equation}
H_\alpha=\alpha\sum_c\left(e_c-\frac{\gamma}{\alpha}\binom{n_c}{2}\right).
\end{equation}
Hence the optimal partitions at resolution $\gamma$ with weights scaled by $\alpha$ equal those at resolution $\gamma/\alpha$ with the original weights.
This identity agrees with the \texttt{leidenalg} maintainer's reply to the issue reported in the thesis~\cite{Arjomandi_2024}; it was also checked numerically on small graphs~\cite{scale_repo}. It holds for pure scaling, not for adding a constant.
For a weighting with mean edge weight $\bar w$, $\gamma/\bar w$ is called the \emph{effective resolution}: the resolution at which unit-mean weights would face the same penalty.

\subsection{Implementation}
\label{sec:impl}
Thesis runs used the \texttt{leidenalg} package on a server with 128~GB of memory and 128 CPUs; a typical clustering took about ten minutes~\cite{Arjomandi_2024}. The clustering script sets the random seed to 4311 by default, and no record of repeated runs with other seeds was found.

New runs used a laptop with 15~GB of memory. The \texttt{leidenalg} baseline exceeded a 10~GB memory cap there, so the new runs use the Leiden implementation in igraph~\cite{igraph} (\texttt{community\_leiden}, CPM objective, two iterations). Its CPM quality is $\sum_c (e_c-\gamma n_c^2/2)$ up to a positive factor, which differs from Eq.~(2) by the constant $\gamma N/2$ for every partition of the $N$ nodes, so both have the same optimal partitions; this was checked numerically~\cite{scale_repo}. Weights are attached to edges in the row order of the feature table, from which the graph is built. A clustering with two iterations takes 11--34 minutes. Ten iterations instead of two (three settings, Sec.~\ref{sec:scaletest}) change Cov1 by at most 0.3 points and the CPM score by under 1\%; a run iterated until the partition stopped changing did not finish within 2.4 hours and was stopped. With seed 4311 and $\gamma=0.05$, the unweighted run gives Cov1 75.54\% and a CPM score of $6.06\times10^{6}$, against 75.9\% and $6\times10^{6}$ in the thesis, and a repeated run gives an identical partition summary. Script, configuration, and per-run outputs are public~\cite{scale_repo}.

\section{Experiments}
\label{sec:results}
This section describes the experimental setup and restates the experiments of the thesis~\cite{Arjomandi_2024}. The unweighted runs serve as the baseline. The weighted runs were made before the fix of Section~\ref{sec:align} and are summarized only as far as the comparison in Table~\ref{tab:align} needs; the full sweeps are in the thesis.

\subsection{Resolution selection}
\label{sec:gammasel}
The thesis clustered the unweighted CENM with $\gamma\in\{10^{-4},10^{-3},10^{-2},0.05,0.25,0.5,0.75,0.95\}$ (Table~\ref{tab:gamma}).
Coverage falls and the number of singletons rises as $\gamma$ grows.
The number of clusters with more than ten nodes peaks at $\gamma=0.05$, so $\gopt=0.05$ was fixed for all weighted runs. At $\gopt$, Cov1 is 75.9\% and the CPM score is $6\times10^{6}$.
The grid was rerun (new runs, seed 4311; last two columns of Table~\ref{tab:gamma}). Cov1 agrees with the thesis within 0.5 points except at $\gamma=0.25$ (23.6\% against 26.1\%), and on the thesis grid the number of clusters with more than ten nodes again peaks at $\gamma=0.05$ (317,595). An extra run at $\gamma=0.075$ gives slightly more (323,951), so the peak lies between 0.05 and 0.1.
The unweighted baseline at $\gopt$ has Cov10 $70.74\pm0.07\%$ (five seeds).
Cov10 falls from 60.6\% at $\gamma=0.075$ to 5.4\% at $\gamma=0.1$, while Cov1 falls smoothly (67.6\% to 54.2\%). The reason follows from $H$: a node with a single edge into a cluster of $n_c$ nodes changes $H$ by $1-\gamma n_c$, so sparse, tree-like clusters stop growing once $n_c\ge 1/\gamma$, at 10 nodes for $\gamma=0.1$ and 14 for $\gamma=0.075$. The cluster sizes confirm this: at $\gamma=0.1$, 301,044 clusters have exactly 10 nodes and 3,160 have 11; at $\gamma=0.075$, 150,041 have 14 nodes and 335 have 15.

\begin{table}[t]
\centering
\caption{Unweighted Leiden-CPM on CENM for varying $\gamma$, as reported in~\cite{Arjomandi_2024}, Table~3.1. Cov1: \% nodes in non-singleton clusters; \#Clus: non-singleton clusters; \#Sing: singletons; CPM: CPM score as printed in the thesis. New: Cov1 and the number of clusters with more than ten nodes (\#$>$10) from new runs (igraph, seed 4311). Unlike the weighted runs, which are scored at $\gamma=0.05$, the thesis code scores each row at its own $\gamma$, so the CPM values are not comparable across rows; this also explains why the value at $\gamma=0.5$ exceeds the one at $\gamma=0.25$. The thesis shows \#$>$10, which sets $\gopt$, only as a plot (Fig.~3.2).}
\label{tab:gamma}
\footnotesize
\setlength{\tabcolsep}{2.5pt}
\begin{tabular}{rrrrrrrr}
\toprule
 & \multicolumn{5}{c}{Thesis} & \multicolumn{2}{c}{New}\\
\cmidrule(lr){2-6}\cmidrule(lr){7-8}
$\gamma$ & Cov1 & \#Clus & \#Sing & Max & CPM & Cov1 & \#$>$10\\
\midrule
$10^{-4}$ & 93.6 & 6,074 & 477,995 & 56,437 & $2.84\times10^{7}$ & 93.19 & 5,883\\
$10^{-3}$ & 91.0 & 37,464 & 673,277 & 17,949 & $2.1\times10^{7}$ & 90.58 & 37,241\\
$10^{-2}$ & 87.5 & 180,361 & 930,049 & 2,545 & $1.22\times10^{7}$ & 87.39 & 177,268\\
\textbf{0.05} & \textbf{75.9} & \textbf{357,527} & \textbf{1,793,229} & \textbf{543} & $\mathbf{6\times10^{6}}$ & \textbf{75.54} & \textbf{317,595}\\
0.25 & 26.1 & 378,047 & 5,501,025 & 142 & $3.56\times10^{5}$ & 23.56 & 11,386\\
0.50 & 13.4 & 331,284 & 6,445,338 & 67 & $6\times10^{5}$ & 13.19 & 6,764\\
0.75 & 12.3 & 369,132 & 6,532,666 & 33 & $2.1\times10^{5}$ & 12.31 & 1,098\\
0.95 & 12.2 & 374,670 & 6,535,729 & 16 & $4\times10^{4}$ & 12.27 & 53\\
\bottomrule
\end{tabular}
\end{table}

\subsection{Weighted runs}
\label{sec:thesisw}
The thesis first used \emph{intuitive} weights: integer importances $1,\dots,7$ (year~1, MeSH~2, co-citation Jaccard~3, co-citation count~4, coupling Jaccard~5, coupling count~6, abstract~7), normalized to sum to one. At $\gopt$, Cov1 dropped from 75.9\% to 31.4\% and the CPM score from $6\times10^{6}$ to $3.26\times10^{6}$~\cite{Arjomandi_2024}.
It then swept the topological features, giving one of them the weight $w\in\{0,\tfrac18,\dots,\tfrac78\}$ and the other three $(1-w)/3$ each, with zero semantic weights. With these weights in $[0,1]$, Cov1 stayed below about 14\% in every setting (thesis plots).
After the nonzero weights were mapped linearly onto $[1,10]$ and zero weights set to $1$, Cov1 ranged from 75.0\% to 76.0\% and Cov10 from 68.3\% to 71.8\%, close to the unweighted baseline; the equal-weight setting gave 75.1\% and 68.6\%. The thesis reports that some rescaled runs did not finish, one after 13 days; the available logs show memory-allocation errors instead, and the run could not be identified, so this is treated as unconfirmed.

Two properties of these weights do not depend on which edge each weight is attached to. With only topological features, 47\% of the edges have weight zero (30,382,764 of 57,512,086 are nonzero), because these citing pairs share neither references nor co-citations; with equal topological weights the smallest nonzero weight is $4.6\times10^{-5}$ (\cite{Arjomandi_2024}, Table~4.3). And the $[1,10]$ map gives weight exactly $1$ to 47.2\% of the edges in the equal-weight setting, with median 1.018~\cite{Arjomandi_A}, so the rescaled weights are close to uniform.

\begin{table*}[t]
\centering
\caption{New runs with aligned weights, $\gamma=0.05$. $\bar w$: mean edge weight; $0.05/\bar w$: effective resolution of the raw weights. Raw: seed 4311. Divided by $\bar w$: mean $\pm$ sd over seeds 4311, 1, 2; baseline: seeds 4311, 1--4. CPM: thesis CPM score in units of $10^{6}$ (mean over seeds). Coherence over intra-cluster edges of the mean-normalized runs: mean $f_6$, share with $f_5=1$, mean $f_2$ (all edges: 0.788, 0.450, 0.0172). Last two rows: topological equal with zero weights replaced before dividing by $\bar w$; $^{*}$seed 4311 only.}
\label{tab:scaletest}
\footnotesize
\setlength{\tabcolsep}{4pt}
\begin{tabular}{lrrrrrrrrrrr}
\toprule
 & & & \multicolumn{3}{c}{Raw weights} & \multicolumn{3}{c}{Divided by $\bar w$} & \multicolumn{3}{c}{Coherence (divided by $\bar w$)}\\
\cmidrule(lr){4-6}\cmidrule(lr){7-9}\cmidrule(lr){10-12}
Weighting & $\bar w$ & $0.05/\bar w$ & Cov1 & Cov10 & CPM & Cov1 & Cov10 & CPM & $f_6$ & $f_5=1$ & $f_2$\\
\midrule
Unweighted & 1 & 0.05 & -- & -- & -- & $75.54\pm0.03$ & $70.74\pm0.07$ & 6.06 & 0.800 & 0.513 & 0.0317\\
Intuitive & 0.2261 & 0.22 & 26.88 & 6.05 & 3.79 & $68.98\pm0.01$ & $65.40\pm0.01$ & 6.29 & 0.812 & 0.596 & 0.0363\\
Topological equal & 0.0081 & 6.20 & 4.73 & 0.13 & 0.38 & $30.75\pm0.01$ & $26.80\pm0.01$ & 4.59 & 0.821 & 0.570 & 0.0523\\
Coupling Jaccard 0.875 & 0.0160 & 3.13 & 8.63 & 0.49 & 0.77 & $32.69\pm0.02$ & $28.48\pm0.02$ & 3.69 & 0.821 & 0.585 & 0.0580\\
Co-citation count 0.875 & 0.0015 & 32.6 & 0.07 & 0.00 & 0.004 & $28.86\pm0.00$ & $24.56\pm0.01$ & 4.70 & 0.822 & 0.570 & 0.0515\\
\midrule
Topological equal, $[1,10]$ & -- & -- & -- & -- & -- & $66.39\pm0.00$ & $63.60\pm0.00$ & 6.33 & 0.808 & 0.536 & 0.0387\\
\quad zeros $\to$ smallest positive & -- & -- & -- & -- & -- & $30.73\pm0.01$ & $26.78\pm0.01$ & 4.60 & 0.821 & 0.570 & 0.0524\\
\quad zeros $\to$ mean positive$^{*}$ & -- & -- & -- & -- & -- & 63.62 & 60.40 & 6.01 & 0.807 & 0.540 & 0.0408\\
\bottomrule
\end{tabular}
\end{table*}

\section{Results}
\label{sec:new}
All runs in this section are new and were not part of the thesis. They use CENM, $\gamma=0.05$ unless stated otherwise, the feature scaling of Section~\ref{sec:method}, and the implementation of Section~\ref{sec:impl}.
Four weightings are used: the intuitive weights of Section~\ref{sec:thesisw}; the four topological features at $1/4$ each (\emph{topological equal}); and coupling Jaccard or co-citation count at $0.875$ with $0.125/3$ for each other topological feature, the end points of two thesis sweeps.

\subsection{Correction of the weight--edge assignment}
\label{sec:align}
The thesis builds the igraph graph from an edge table through networkx (\texttt{from\_pandas\_edgelist}, relabeling, \texttt{from\_networkx}), and passes the weight column of a separate table, ordered by an edge identifier, to \texttt{leidenalg} by position. Nothing maps the weights to the graph's edge order.
The thesis functions, copied unchanged, were applied to random edge tables: igraph edge $k$ was table row $k$ for only 2 of 10,000 edges, because networkx lists edges grouped by endpoint. Applied to the row order of the local feature table, 3 of 57,512,086 edges keep their position. Unless the thesis tables happened to be in networkx order, every weight was attached to some other edge.

To test this, two thesis settings were rerun with three assignments: aligned; the thesis assignment, assuming the thesis tables had the local row order; and a random permutation, which needs no assumption about the order (Table~\ref{tab:align}).
Both misaligned assignments reproduce the thesis: for the intuitive weights, the CPM score is $3.26\times10^{6}$ (thesis $3.26\times10^{6}$) and the largest cluster has 158 nodes (thesis 156); for the rescaled equal weights, Cov1 is 75.2\% (thesis 75.1\%). The aligned runs differ clearly from the thesis. The two misaligned variants agree with each other within 0.04 percentage points, so the result does not depend on the unknown row order. For raw topological equal weights, the thesis gives only plotted values (Cov1 below about 14\%); both the aligned run (4.73\%) and the thesis assignment (9.43\%) fall in that range, so this setting does not discriminate. The remaining differences from the thesis are of the size expected from the change of Leiden implementation.

This was an error in the thesis code. It affects all weighted clustering results of the thesis (the intuitive weights, the topological sweeps, and the $[1,10]$ rescaled runs), and not the unweighted runs, the graph construction, or the edge features; an erratum is in the README of the code repository~\cite{scale_repo}. All new runs attach each weight to its own edge.
Whereas the thesis reported that $[1,10]$ rescaling keeps coverage near the baseline, the updated method shows that this is what near-uniform weights unrelated to the edges give; with aligned weights, the same map lowers Cov1 by 9 points.

\begin{table}[t]
\centering
\caption{Thesis values and new runs (seed 4311, $\gamma=0.05$) with aligned weights, the thesis weight assignment (misaligned), and randomly permuted weights. CPM: thesis CPM score in units of $10^{6}$. Thesis values from~\cite{Arjomandi_2024} (Sec.~4.1; Tables~4.4--4.7, strength $0.25$); --: not reported.}
\label{tab:align}
\footnotesize
\setlength{\tabcolsep}{3pt}
\begin{tabular}{lrrrrr}
\toprule
 & Cov1 & Cov10 & \#Clus & \#Sing & CPM\\
\midrule
\multicolumn{6}{l}{\emph{Intuitive weights, raw}}\\
Thesis & 31.4 & -- & 397,461 & 5,110,353 & 3.26\\
Aligned & 26.88 & 6.05 & 304,836 & 5,444,156 & 3.79\\
Thesis assignment & 32.28 & 3.10 & 401,652 & 5,042,145 & 3.26\\
Permuted & 32.30 & 3.09 & 401,795 & 5,040,699 & 3.26\\
\midrule
\multicolumn{6}{l}{\emph{Topological equal, rescaled to $[1,10]$}}\\
Thesis & 75.1 & 68.6 & -- & 1,854,506 & --\\
Aligned & 66.39 & 63.60 & 259,915 & 2,502,865 & 6.33\\
Thesis assignment & 75.18 & 67.88 & 343,386 & 1,848,141 & 5.49\\
Permuted & 75.14 & 67.90 & 343,008 & 1,851,364 & 5.49\\
\bottomrule
\end{tabular}
\end{table}

\subsection{Weight scale with aligned weights}
\label{sec:scaletest}
Table~\ref{tab:scaletest} gives, for each weighting, the mean weight $\bar w$ and the effective resolution $0.05/\bar w$, and clusters it twice at $\gamma=0.05$: with raw weights, and with weights divided by $\bar w$, which by the identity of Section~\ref{sec:scale} equals raw weights at $\gamma=0.05\,\bar w$. Mean-normalized runs were repeated with three seeds and the baseline with five.

The effective resolutions range from 0.22 to 32.6, all above $\gopt$. With raw weights, coverage collapses accordingly: Cov1 is 26.9\% for the intuitive weights, close to the unweighted 26.1\% at $\gamma=0.25$ (Table~\ref{tab:gamma}), and below 9\% for the topological weightings.
Dividing by $\bar w$ removes most of this effect for the intuitive weights: Cov1 rises to 69.0\% and Cov10 to 65.4\%, against 75.5\% and 70.7\% unweighted.
For the topological weightings it does not: Cov1 rises only to 29--33\%. These weightings give weight zero to the 47\% of edges without shared references or co-citations, whereas the intuitive weights include $f_5$ and $f_6$ and are positive on nearly every edge.
Lowering $\gamma$ does not fix this: at $\gamma=0.01$, the mean-normalized topological weights reach only 37.2\% Cov1 (Fig.~\ref{fig:coh}), while the unweighted graph reaches 87.4\%.
The zero weights were then replaced before dividing by $\bar w$ (last rows of Table~\ref{tab:scaletest}). Raising them to the smallest positive weight, which is below 1\% of the mean, changes nothing (30.73\% Cov1 over three seeds). Setting them to the mean positive weight, so that these edges count like an average edge, doubles Cov1 to 63.6\% and Cov10 to 60.4\%, close to the $[1,10]$ map (66.4\% and 63.6\%).
So the edges without topological overlap hold together most of the coverage, and a weighting that makes them nearly weightless loses it, whether their weight is exactly zero or merely tiny.

Seed-to-seed standard deviations are at most 0.07 percentage points for Cov1 and Cov10 in every setting, far below the differences between settings.
The CPM score of two weighted runs (intuitive divided by $\bar w$, $6.29\times10^{6}$; aligned $[1,10]$, $6.33\times10^{6}$) exceeds the baseline's $6.06\times10^{6}$, although the baseline optimizes this score directly. With ten iterations instead of two, the three scores rise only slightly ($6.32$, $6.37$, and $6.11\times10^{6}$) and keep their order. So Leiden stops at local optima, and the weights can steer it to a better one.

\subsection{Coherence of the clusters}
\label{sec:coherence}
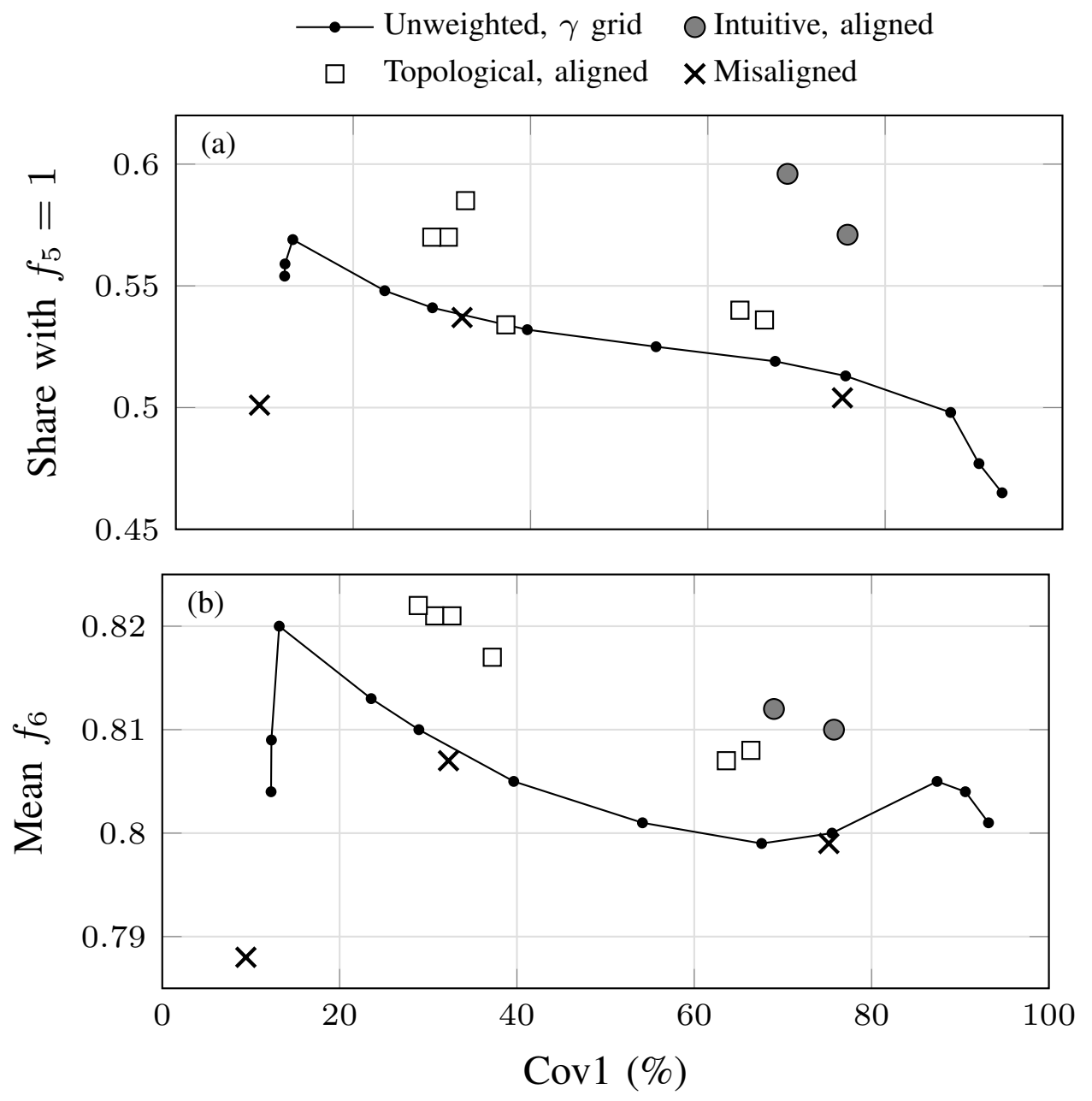
\begin{figure}[t]
\centering
\begin{tikzpicture}
\begin{axis}[scale only axis, width=0.8\columnwidth, height=3.2cm, xmin=0, xmax=100, ymin=0.45, ymax=0.62,
  ylabel={Share with $f_5=1$}, xticklabels={},
  xticklabel style={font=\scriptsize}, yticklabel style={font=\scriptsize, /pgf/number format/fixed, /pgf/number format/precision=3}, label style={font=\small},
  legend style={font=\scriptsize, at={(0.5,1.02)}, anchor=south, draw=none, legend columns=2, /tikz/every even column/.append style={column sep=6pt}}, legend cell align=left,
  grid=major, grid style={gray!25}]
\addplot[black, thin, mark=*, mark size=1pt] coordinates {(93.19,0.465)(90.58,0.477)(87.39,0.498)(75.54,0.513)(67.6,0.519)(54.16,0.525)(39.64,0.532)(28.94,0.541)(23.56,0.548)(13.19,0.569)(12.31,0.559)(12.27,0.554)};
\addlegendentry{Unweighted, $\gamma$ grid}
\addplot[only marks, mark=*, mark size=2.2pt, fill=gray] coordinates {(68.99,0.596)(75.76,0.571)};
\addlegendentry{Intuitive, aligned}
\addplot[only marks, mark=square*, mark size=1.9pt, fill=white] coordinates {(66.39,0.536)(30.73,0.57)(37.22,0.534)(63.62,0.54)(32.67,0.585)(28.87,0.57)};
\addlegendentry{Topological, aligned}
\addplot[only marks, mark=x, mark size=3pt, thick] coordinates {(32.28,0.537)(75.18,0.504)(9.43,0.501)};
\addlegendentry{Misaligned}
\node[font=\scriptsize, anchor=north west] at (axis cs:1,0.62) {(a)};
\end{axis}
\end{tikzpicture}

\vspace{1mm}
\begin{tikzpicture}
\begin{axis}[scale only axis, width=0.8\columnwidth, height=3.2cm, xmin=0, xmax=100, ymin=0.785, ymax=0.825,
  ylabel={Mean $f_6$}, xlabel={Cov1 (\%)},
  xticklabel style={font=\scriptsize}, yticklabel style={font=\scriptsize, /pgf/number format/fixed, /pgf/number format/precision=3}, label style={font=\small},
  legend style={font=\scriptsize, at={(0.5,1.02)}, anchor=south, draw=none, legend columns=2, /tikz/every even column/.append style={column sep=6pt}}, legend cell align=left,
  grid=major, grid style={gray!25}]
\addplot[black, thin, mark=*, mark size=1pt] coordinates {(93.19,0.801)(90.58,0.804)(87.39,0.805)(75.54,0.8)(67.6,0.799)(54.16,0.801)(39.64,0.805)(28.94,0.81)(23.56,0.813)(13.19,0.82)(12.31,0.809)(12.27,0.804)};

\addplot[only marks, mark=*, mark size=2.2pt, fill=gray] coordinates {(68.99,0.812)(75.76,0.81)};

\addplot[only marks, mark=square*, mark size=1.9pt, fill=white] coordinates {(66.39,0.808)(30.73,0.821)(37.22,0.817)(63.62,0.807)(32.67,0.821)(28.87,0.822)};

\addplot[only marks, mark=x, mark size=3pt, thick] coordinates {(32.28,0.807)(75.18,0.799)(9.43,0.788)};

\node[font=\scriptsize, anchor=north west] at (axis cs:1,0.825) {(b)};
\end{axis}
\end{tikzpicture}
\caption{Coherence of intra-cluster edges against coverage (new runs, seed 4311). Line: unweighted runs over $\gamma$ from $10^{-4}$ (right) to $0.95$ (left). Filled circles: intuitive weights divided by $\bar w$ at $\gamma=0.05$ and $0.035$ (these weights include $f_5$ and $f_6$). Squares: topological weightings with aligned weights (divided by $\bar w$, at $\gamma=0.05$ and $0.01$, with zero weights set to the mean, and $[1,10]$), which do not use $f_5$ or $f_6$. Crosses: misaligned runs of Table~\ref{tab:align} and the raw topological run. Over all edges, the share is 0.450 and the mean $f_6$ 0.788.}
\label{fig:coh}
\end{figure}

Every aligned weighting raises all three coherence measures over the baseline (Table~\ref{tab:scaletest}), whereas the misaligned $[1,10]$ run stays close to it (0.799, 0.504, and 0.0310, against 0.800, 0.513, and 0.0317), as weights unrelated to the edges should.
Coherence also rises when coverage falls, because a stricter partition keeps only the tightest pairs together. Fig.~\ref{fig:coh} therefore plots coherence against Cov1. The unweighted runs over the $\gamma$ grid form a reference curve; a run above it is more coherent than unweighted clustering with the same coverage.

All aligned weighted runs lie above the curve for $f_6$, and all but one lie above it for $f_5$ (the topological weights at $\gamma=0.01$ lie on it); all misaligned runs lie on or below it.
At Cov1 near 67--69\%, the share of intra-cluster edges with $f_5=1$ is 0.519 for the unweighted run at $\gamma=0.075$, against 0.596 for the intuitive weights and 0.536 for the rescaled topological weights. At Cov1 near 30\%, it is 0.541 unweighted ($\gamma=0.2$) against 0.570--0.585 for the mean-normalized topological weightings. The most coherent unweighted run ($\gamma=0.5$, 0.569) clusters only 13\% of the nodes.
With the intuitive weights at $\gamma=0.035$, Cov1 (75.76\%) and the CPM score ($6.08\times10^{6}$, one seed) match the baseline's, Cov10 rises from 70.71\% to 73.09\%, and the share of edges inside clusters from 0.164 to 0.196.

The gains are modest. At matched coverage, mean $f_6$ rises by 0.009--0.013, or 12--17\% of its interquartile range on CENM (0.076~\cite{Arjomandi_A}). The share with $f_5=1$ moves more, by up to 0.08, but the largest gains belong to the intuitive weights, which include $f_5$ and $f_6$, so they are partly circular.
The topological weightings do not use $f_5$ or $f_6$. Part of their year gain is still expected, because coupling overlap and year proximity go together: the share of edges within five years rises from 34.6\% for edges without topological overlap to 70.9\% in the top coupling-Jaccard quintile~\cite{Arjomandi_A}. The gains are therefore real gains on features outside the weight, but they partly reflect this correlation.
The weighted runs also keep more edges inside clusters at equal coverage (0.161 against 0.068 at Cov1 near 30\%), so weighting changes which articles are grouped together, not only how many are clustered.

\section{Discussion}
\label{sec:discussion}
\emph{What the results show.}
First, whereas the thesis reported a sharp loss of coverage for weights in $[0,1]$ and near-baseline coverage after rescaling to $[1,10]$, the updated method shows that this pattern follows from the weight distribution alone, not from the features.
Second, with aligned weights, weight scale explains most of the loss of coverage: raw weights with mean below one raise the effective resolution, and dividing by the mean restores most of the coverage when nearly all weights are positive.
Third, where 47\% of edges have weight zero, neither normalization nor a lower $\gamma$ is enough, and coverage stays below half of the baseline's; these edges hold most of the coverage, and giving them an average weight restores it.
Fourth, at equal coverage, aligned weighting makes the clusters somewhat more coherent, also on features that are not part of the weight, and keeps more edges inside clusters. With the intuitive weights at $\gamma=0.035$, coverage equals the baseline's and Cov10 is higher.

\emph{Practical guidance.}
(i) Attach weights to edges by key or as an edge attribute, not by position, and check the result, for example by recomputing one feature from the clustered graph's edge list. (ii) Normalize weights to mean one, or report the effective resolution $\gamma/\bar w$. (iii) Count zero- and near-zero-weight edges: they contribute almost nothing to the objective, yet in CENM they hold together most of the coverage. (iv) Choose $\gamma$ for the weighted graph, or compare runs at equal coverage, rather than reusing the unweighted $\gopt$. (v) Repeat runs with a few seeds; at this scale each costs under an hour on a laptop, and the observed variation was below 0.1 points.

\emph{Limitations.}
(1) One graph; most weighted runs use the unweighted $\gopt$, and only two lower values of $\gamma$ were tried for weighted graphs.
(2) The new runs use igraph's Leiden instead of \texttt{leidenalg}; the unweighted baseline agrees with the thesis within 0.4 points of Cov1 at $\gopt$, and within 0.5 points elsewhere on the grid except at $\gamma=0.25$ (2.5 points).
(3) The coherence measures are internal and partly circular for weightings that include the measured feature; they are not a ground-truth evaluation, and following Peel et al.~\cite{Peel_2017}, metadata and topology need not agree.
(4) The semantic features were not swept, and the weighting space was not searched exhaustively.
(5) The misaligned-assignment test assumes the thesis tables had the local row order; the random-permutation test does not, and gives the same result.
(6) No comparison with the evaluation of Ahlgren et al.~\cite{Ahlgren_2020}.
(7) The runs at other $\gamma$ values and the zero-replacement run at the mean use one seed; elsewhere the seed spread was below 0.1 points.

\section{Conclusion}
A 7.4-million-node PubMed citation graph was clustered with Leiden-CPM, using edge weights built from seven similarity features.
With the method updated to attach each weight to its own edge, weight scale explains most of the loss of coverage, and edges without topological overlap, when weighted near zero, explain most of the rest. At equal coverage, weighting makes clusters more coherent, also on features outside the weight.
Next steps are to choose $\gamma$ systematically for each normalized weighting and to evaluate against external labels.

\section*{Data and Code Availability}
Parsed PubMed metadata are available at the Illinois Data Bank~\cite{Arjomandi_2024_data}, and the parsing code is at \url{https://github.com/illinois-or-research-analytics/pubmed_etl}.
The thesis code for graph construction, features, weighting, and clustering is available as a cleaned copy~\cite{thesis_pipeline_repo} of the original repository~\cite{hm31_graphs_repo}; its clustering step attaches weights to edges by position, which the new runs fix (Section~\ref{sec:align}), and the README of \texttt{pubmed-meta-etl} holds the corresponding erratum.
The script, configuration, and per-run outputs (aggregate statistics only) of the new runs, and the edge-order check of Section~\ref{sec:align}, are at~\cite{scale_repo}.
The CENM graph and its weighted edge list are not publicly released, because they derive from the research group's network data.

\section*{Acknowledgment}
The author thanks George Chacko for reading the manuscript and offering suggestions.

Use of generative AI: Claude (Anthropic) was used to draft and edit the text of all sections, to write the \LaTeX{} source, including the tables and Fig.~\ref{fig:coh}, and to write the scripts for the numerical check in Section~\ref{sec:scale}, the edge-order check, and the new clustering runs in Section~\ref{sec:new}. All numbers come from the thesis, the processing logs, or the outputs of these scripts, and were checked by the author, who reviewed and validated all content and takes full responsibility for it.

\bibliographystyle{IEEEtran}
\bibliography{refs}

\end{document}